# Development of a One Dollar Blood Pressure Monitor

Yinan Xuan, Ava J. Fascetti, Colin Barry, and Edward J. Wang

*Abstract—* BPClip is an ultra-low-cost cuffless blood pressure monitor. As a universal smartphone attachment, BPClip leverages the computational imaging power of smartphones to perform oscillometry based blood pressure measurements. This paper examines different design considerations in BPClip's development. The cost and accuracy of blood pressure measurements are the central design goals. Both of these requirements are achieved with the initial prototype that achieves a $0.80 USD material cost and a mean absolute error of 8.72 and 5.49 mmHg for systolic and diastolic blood pressure, respectively. Since a main motivator to develop BPClip is making blood pressure monitoring more accessible, usability is also central to the design. User studies were conducted throughout the design process to inform the most intuitive and accessible design features. In this paper, we demystify the design process to share effective design practices with future developers working towards expanding health monitoring access beyond traditional clinical settings.

*Clinical Relevance—* This solution can provide wide access to affordable blood pressure monitoring. As an ultra-low-cost monitor, BPClip can be deployed into vulnerable communities at little or no cost, vastly increasing screening capabilities.

## I. INTRODUCTION

Hypertension is the leading preventable cause of premature death and disability globally, claiming almost eight million lives each year. By 2025, it is projected to increase by 60% to affect 1.6 billion adults [1]. In addition to elevating the risk of cardiovascular diseases like stroke, heart disease, chronic kidney disease, and end-stage kidney disease, hypertension has been shown to accelerate cognitive decline in middle-aged and older adults. An estimated 46% of adults living with hypertension, over 500 million people, are unaware and undiagnosed [2].

Traditionally, the diagnosis and management of hypertension have been dependent on blood pressure (BP) measurements taken during office visits. However, office BP measurements are subject to white-coat hypertension, are taken less frequently, and may miss temporal abnormalities. As such, there is a need for self-measured oscillometric BP monitoring as it may improve the diagnosis and management of hypertension [3]. Self-measured devices can facilitate wide scale screening capabilities if deployed effectively.

Self-measured BP monitoring can also make a significant impact in low-income communities where barriers to healthcare can lead to increased risk for chronic diseases like hypertension [4]. These communities include older adults in rural areas, pregnant women especially those from low-income backgrounds, refugees, and African Americans. Research has shown that one way to more effectively reach these vulnerable populations is to deploy health monitoring directly into their communities. For example, barbershop-based health promotion has been successful in reaching thousands of men nationwide [5].

[1]*Research supported by NIH MassAITC P30AG073107, Google Research Scholar Award, and UCSD GEM Prize.

E.J.W. is an Assistant Professor at UCSD, La Jolla, CA 92037 USA,; e-mail: ejaywang@ ucsd.edu).

We introduce **BPClip**, an ultra-low-cost universal smartphone attachment that enables smartphones to measure blood pressure at-home, in barbershops, or anywhere. This solution can provide wide access to affordable BP monitoring as it has no electronic components, does not require calibration with a cuff or prior BP measurements, and is compatible with any smartphone featuring a low-resolution (2MP) camera and a light source. Unlike prior works that enable phone-based measurements with external sensors [6], our innovation involves employing computational imaging using a smartphone's camera with a cheap plastic clip. We envision that BPClip can be delivered at scale for mass screening efforts at community events, in organizations, or even through the mail.

## II. BPClip

### A. BPClip System

The BPClip is an ultra-low-cost cuffless blood pressure monitor that leverages the hardware and computational power of modern smartphones. To use BPClip, the user presses the clip with their index finger and holds pressure at discrete force levels as prompted by a smartphone app. The phone's flashlight illuminates the fingertip through a pinhole on the clip and the phone's camera captures the brightness of the pinhole projection.

To measure blood pressure, BPClip performs oscillometry at the fingertip. Oscillometry, which is a measurement of the arterial pulse amplitude as a function of applied pressure, requires the ability to apply pressure on the artery, sense how much pressure is applied on the artery, and measure the blood volume oscillation. To apply pressure on the artery (transverse palmar arch artery), the user presses down on the device with their index finger. The movable pinhole components rest on a spring-loaded mechanism so the more the user presses down, the closer the pinhole moves to the camera, increasing its projection size. Therefore, the pressure on the artery is encoded by the size of the pinhole projection from the camera's perspective. The blood volume oscillation is detected by the camera from the brightness of the pinhole projection based on the photoplethysmograph (PPG) effect. With this data at different force levels, blood pressure is then calculated with a linear regression model.

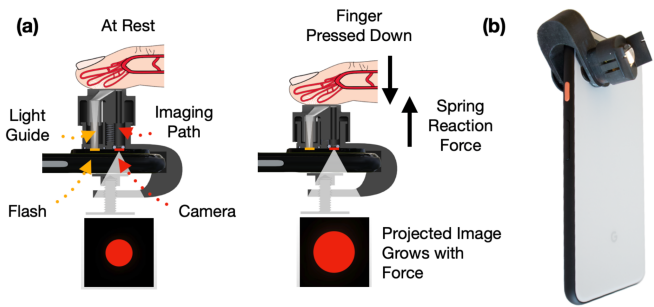

**Figure 1.** (a) BPClip mechanism concept. (b) BPClip attached.

## B. BP Monitoring Performance

The BPClip was tested in a human subjects study of 24 people. Measurements from the system were compared to a standard arm cuff device for reference. Subjects had systolic blood pressures ranging from 88 to 157 mmHg and diastolic blood pressures ranging from 57 to 97 mmHg. The BPClip measurements achieved a mean absolute error of 8.72 mmHg and 5.49 mmHg for systolic and diastolic blood pressure, respectively.

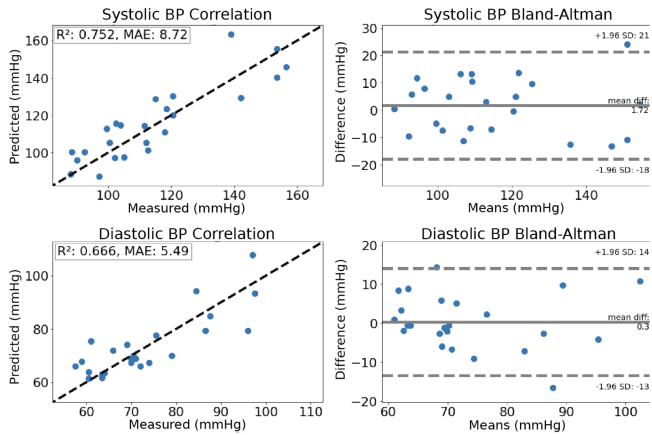

Figure 2. BPClip performance. (N=24)

There was no significant difference in error among the three ethnic groups included in the study. Subgroup analysis on the BPClip accuracy showed mean absolute error for systolic blood pressure for Asian, Hispanic, and White groups were $7.7 \pm 4.2$, $8.6 \pm 5.2$, and $10.2 \pm 6.9$ mmHg, respectively. For diastolic blood pressure, the errors were $4.5 \pm 4.6$, $9.5 \pm 6.4$, and $6.5 \pm 2.9$ mmHg, respectively [7].

## C. Cost of BPClip

One major design feature of the BPClip is its ability to perform accurate BP measurements while being ultra-low-cost. This enables BPClip to be provided at little to no cost to vulnerable communities by nonprofits or public health initiatives to reduce costly medical expenditures from the high prevalence of cardiac conditions. By harnessing the smartphone's imaging system for computational power, the clip only requires a spring and other widely available mechanical components. This keeps the material cost under $1USD. In a cost projection study of 1000 units, the cost per clip of $0.80 breaks down to the spring ($0.03), the acrylic cover ($0.001), two O-rings ($0.04), the anti-slip pads ($0.03), the metal rod and tube ($0.15), and the resin 3D printed material ($0.55) [6]. When mass produced, per component costs will decrease further as they can be purchased at bulk and the cost of the 3D printed material will also decrease as injection molding can be used. However, even at low production, BPClip costs more than 10 times less than any blood pressure monitor on the market.

## III. SYSTEM DESIGN CONSIDERATIONS

### A. Measuring Pressure on a Smartphone without a Pressure Sensor

Relying on the smartphone's existing sensors to perform BP measurements is a central requirement for the system. This allows the clip to be fully mechanical and removes the reliance on external electronics. Multiple methods were evaluated in the pursuit of measuring pressure without a pressure sensor. The two methods described here are the capacitive screen with conductive polymer mechanism and the movable pinhole mechanism.

The capacitive screen with conductive polymer design utilized the 3D printable, conductive, flexible thermoplastic polyurethane (TPU). The finger would compress the TPU into the screen of the phone. Due to its conductive properties, the contact area of the TPU is then read by the screen to calculate force exerted by the finger. The biggest challenge with this method, which was ultimately the reason it was not pursued further, was the lack of direct software access to the screen data. Other complications included the fact that the force to area relationship would need to be calibrated for the 3D printed TPU, and that it was unclear how to get the simultaneous PPG reading of the finger for the oscillometric measurement.

The movable pinhole method is the current mechanism in BPClip. This design does not require touch screen software support as the data collection is functional across devices as it only uses the camera for sensing. While simpler in pressure calculation, this method relies on an additional mechanism in the design to induce the variable pressure that facilitates the force-distance relationship used for pressure calculations.

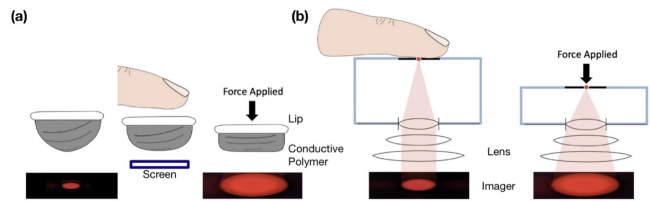

Figure 3. Concepts for measuring pressure without a pressure sensor. (a) Capacitive screen with conductive polymer. The TPU flattens, increasing its area with the screen with greater force. (b) Movable pinhole. Pressure is encoded in the circle size based on the force-distance relationship.

### B. Mechanisms for Inducing Variable Pressure

Several mechanisms were considered for inducing variable pressure. These include the use of a compressible urethane foam, 3D printing flexible materials, 3D printing with a cantilever, and a metal spring.

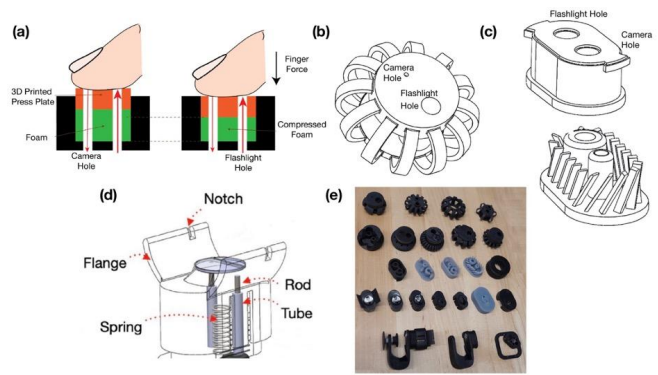

Figure 4. Inducing Variable Pressure. (a) Compressible Urethane Foam. (b) 3D Printing Rubber. (c) 3D Printing with a Cantilever. (d) Metal Spring. (e) Iterative Design Process.

A vision at the beginning of the design process was to make an entirely 3D printable system. This would keep the manufacturing process cheap and flexible in shape and size. However, the most important constraints remain the functionality of the system and the ability of the device to provide accurate BP measurements. To this end, the final mechanism for inducing variable pressure must meet calibration, tolerance, and fatigue requirements so as to not compromise its reliability in producing accurate BP measurements over time. A spring mechanism able to deflect and recover in a known manner is needed to meet these requirements.

Compressible urethane foam is a 3D printable, pressable material that meets the requirements in its ability to be flexible in shape and low in cost. However, experimental trials revealed hysteresis during data collection. This lack of uniformity led to other methods being explored.

3D printing rubber was next evaluated because it also met the low cost and easy assembly objectives. Like the compressible urethane foam, the goal was to be able to "press" the material with varying force to achieve distinct pressure levels. Manufacturing the rubber proved difficult, though, as the rubber was not consistent from batch to batch, which would lead to variabilities when mass produced. Additionally, the rubber tended to deform isotropically, rather than just in the vertical direction, which would lead to inaccurate calculations.

A 3D printed cantilever mechanism sought to resolve the prior problem by restricting random movement in the x-y direction. While meeting many of the objectives, the cantilever system failed fatigue requirements. As the plastic resin was a rigid material, it experienced brittle fracture even within short-term lab experiments. Additionally, the friction experienced by the cantilever varied over time, resulting in inconsistent measurements.

The final mechanism evaluated was to incorporate a metal spring with a 3D printed body, which was used in the end design. While compromising the fully 3D printable objective, it proved to be the only option that met the required constraints. The spring does not require nonlinear calibration, has a defined distance-force relationship that does not vary over time, and has high tolerance. While the metal spring is not 3D printable, it is still low cost at $0.03. Selection of a metal spring to induce variable pressure ensured a reliable z force measurement while meeting the general low-cost requirement, despite not being 3D printable.

*C. Light Transduction*

The last requirement to perform oscillometry is the ability to measure the blood volume oscillation using the brightness of the pinhole projection detected by the smartphone's camera. Maximizing light transfer and minimizing light leakage are central to optimizing for this measurement. Additionally, the camera must only detect brightness from the pinhole projection, not from ambient light or the flashlight itself. Several design decisions were made with these goals in mind.

To ensure the camera only measures brightness from the pinhole, a light guide was incorporated into the design. This separates the flashlight channel from the camera channel. In initial testing without the light guide, the signal was too weak as the flashlight overpowered much of the brightness detected by the camera.

To increase signal quality, maximizing light transfer is crucial. As the clip is made of a black resin material, significant light is lost to absorption in the light guide. To circumvent this, chrome spray paint was applied to the interior of the light guide channel. By coating the interior channel with chrome paint, we maximize the photon transfer from the flashlight to the finger. Additionally, to minimize the amount of light that leaks out of the imaging channel, o-rings are included to ensure a proper seal and restrict ambient light from affecting the brightness of the pinhole projection.

One challenge that arises with the incorporation of the light guide is that some phone models vary in distance between the camera and the flashlight. As such, we conducted experiments to determine the amount of light power lost in the designed light guide given different flashlight to camera distances. Our experiments showed that with the inclusion of the chrome paint, the current design supports distances up to 16mm [7]. This was deemed satisfactory given a survey of flashlight to camera distances of phones on the market showed that the vast majority of phones have less than 16mm distances [7]. However, as a central design goal for BPClip is accessibility to use the device on any smartphone available, future design work is required to keep BPClip up-to-date as phones in the future may be released with significant design changes.

IV. USABILITY CONSIDERATIONS

An important feature of an accessible health monitoring device is usability. Just as important as the measurement of BP itself, is the ability for users to understand how to properly use the device. As BPClip is a self-measurement system, it's important to not add any barriers to accessing the technology through complexity or burden in taking measurements. As such, many usability design considerations were explored to make the clip as straightforward and easy to use as possible.

*A. Finger Placement*

Accurate BP readings require proper orientation of the finger in the clip. Therefore, it is necessary to somehow guide the user to place their finger in the correct position.

The first method used was verbally telling users how to orient their finger. Unfortunately, the design was not as intuitive as we initially believed because despite verbal instructions, users still had a hard time following them correctly. This method would also not be sustainable outside of a medium-sized study.

To remove reliance on verbal instructions and to make the design more self-explanatory, we implemented physical restraints. Two flanges on either side of where the finger slides in forces the finger to be in a specific position. In user testing with this design, we encountered significantly less confusion among participants self-measuring their BP.

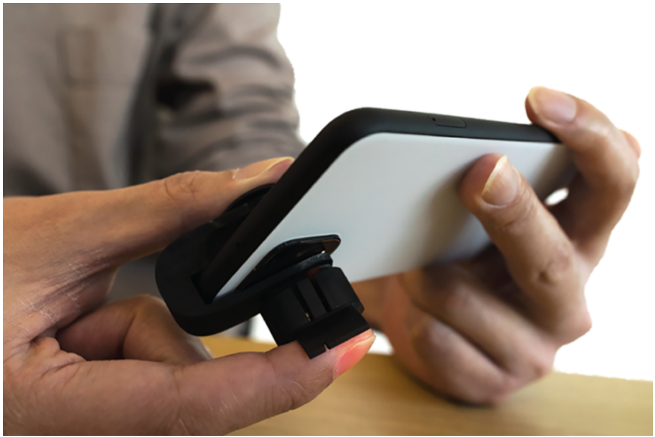

**Figure 5.** Using BPClip. The right index finger slots between the two flanges and the base of the finger aligns with the notch.

*B. Continuous vs Stepwise Pressing*

Another consideration in the design was the method of which the user operates the device. Two options were explored for prompting the user to press the clip at varying force levels. With continuous pressing, the user gradually varies finger force while the camera continuously records data. With stepwise pressing, the user holds their finger at discrete force levels for data collection before moving to the next force level.

Stepwise pressing was selected for the design based on initial user testing. It proved easier for users to lock their finger at discrete force levels for consistent readings than to exhibit the fine control required for steady continuous sensing. Another advantage of stepwise pressing is that the system can take multiple readings at each force level for redundancy. With continuous pressing, real time feedback would be required to prompt the user to fix or redo their pressing if pressure readings were erroneous.

*C. Increasing vs Decreasing Force*

The decision to prompt users for increasing or decreasing force levels was the most subjective design decision described in this paper. Information for this decision was mainly based on user studies and individuals' comfort levels with each method. For some, increasing force was easier, while others preferred decreasing force. Ultimately, increasing the force was selected for the final design for two reasons. First, it is better to fail the test immediately than after the entire data collection process. User studies have shown that people have the most difficulty with the lowest force level because it requires the finest level of control. By increasing the force, users first attempt this difficult level and then can proceed with the rest of the testing if successful. If they need to adjust their grip or otherwise alter their method, they don't need to go through the whole measurement process for no result. Second, when users experimented with decreasing force, oftentimes they would completely let go of the clip on the lowest force level, messing the entire measurement up. Despite some users preferring the decreasing force method, these two reasons warranted increasing force being selected for the design.

## V. Conclusion

The BPClip is a novel ultra-low-cost solution for blood pressure monitoring as it facilitates calibration-free measurement with a smartphone. The system is compatible with any phone that has a flash and a camera. This paper examines different design considerations in the development of BPClip.

System design considerations were evaluated against the goals of keeping costs low and maintaining functionality. Advice to aspiring developers is to question and explore every part of the design that you can within the constraints of the problem. Finding clever solutions may be appealing, but they can't sacrifice the functionality of the system. In BPClip, this is seen in the compromise of a 3D printed spring mechanism in favor of a metal spring.

User centered experiments are a non-negotiable part of the design process. It is impractical to design a system for accessibility and usability without in person experiments with real users and people. Designs must be evaluated in real situations to validate their performance. BPClip relied on user testing throughout the design process to ensure the final device is accommodating to users of various backgrounds.

BPClip applied design principles to prototype ultra-low-cost blood pressure monitors for more accessible screening solutions. This paper aims to demystify the design process and educate future developers who can continue to iterate on accessible health monitoring tools to bring widespread, at-home access to crucial vitals without the reliance on existing clinical infrastructure.